# Developing Augmented Reality based Gaming Model to Teach Ethical Education in Primary Schools

Mohammad Ali
Computer Science and Engineering
United International University
Dhaka, Bangladesh
mali122013@bscse.uiu.ac.bd

*Abstract*— Education sector is adopting new technologies for both teaching and learning pedagogy. Augmented Reality (AR) is a new technology that can be used in the educational pedagogy to enhance the engagement with students. Students interact with AR-based educational material for more visualization and explanation. Therefore, the use of AR in education is becoming more popular. However, most researches narrate the use of AR technologies in the field of English, Maths, Science, Culture, Arts, and History education but the absence of ethical education is visible. In our paper, we design the system and develop an AR-based mobile game model in the field of Ethical education for pre-primary students. Students from pre-primary require more interactive lessons than theoretical concepts. So, we use AR technology to develop a game which offers interactive procedures where students can learn with fun and engage with the context. Finally, we develop a prototype that works with our research objective. We conclude our paper with future works.

## I. Introduction

Technology is growing rapidly and modernizing every sector of our daily life. The education sector is one of them and it has been adopting new technologies to resolve current issues. Digital materials such as audio, video, e-book, and simulations have already been introduced to teach students in an effective way and the result shows that students learn better than traditional methods. But there are some interaction and engagement issues while teaching with those digital materials [1] [2]. Augmented Reality (AR) is a new technology that can be used to resolve those issues. AR creates 3D virtual object, projects it in the real physical world by smart devices like the smartphone with a camera. Therefore, it can be used in educational content with interaction and engagement capabilities [3].

Many researchers have conducted and showed that AR has an effective visualization of real-world events that enhance the capabilities of learning. Researchers built an AR-based educational device that presents realistic replication of earth processes in the field of geoscience education. Researchers prove that it helps students to develop critical thinking and improve engagement in a geoscience classroom [4]. In the paper[5], researchers show visual impacts of temperature on the human inside a house by using AR technology which is a context from the textbook for architecture students. Students use phone-based Augmented Reality technology to change the parameters of a house which reflects energy transformation and affects the temperature inside the room. AR technology can be also useful for students with disabilities, researchers propose an AR-based solution for deaf readers which can enhance their vocabulary learning. The device captures the image and detects the word located above the student's fingertip, an OCR algorithm identifies the meaning of the word and matches the definition from the Sign Language library, and show the video to the student [6]. AR technology is also using for training and other educational purposes, a research work proposes a game-based AR concept to teach users about the library instruction system and proves that users' knowledge can be enhanced and it gives the same results in the same learning performance as conventional library instruction [7]. So, AR technology has lots of benefits in education regardless of the level of students and level of the institution.

In this paper, our target group students are primary level students because primary level students have the best capturing capabilities. AR-based digital books related to Science, Culture, Arts, and History improve engagement and enjoyment during their learning process as well as easy understudying of the subject [8]. AR-based interactive games also have the potential for pre-primary students to increase their logical and mathematical skills [9] [10], students can explore old historical architectures using the game with different AR activities supported by their textbook [11].

Our paper describes how to use AR technologies effectively for pre-primary students to improve their learning experience in the field of ethical education. We develop a mobile-based AR model where students can learn some ethical parts: Conflict Resolution, Ethical Action, Justice, Respect, Responsibility, and Self-esteem in an interactive way. We also develop a prototype on a particular ethical topic: Justice to show how theoretical model can be more interactive with AR technology.

## II. Related Work

A study proposes a framework that intends to help the students build metaphors that correlate body motions and gesture mnemonics with mathematical elements. Therefore, they use an existing novel curriculum, integrate with AR technology to teach mathematics in an interactive way [12]. In another work, researchers take context from children's books to propose a method to integrate AR technology with text, sound, and pop-up cartoon to present different information based on the learning objectives to enhance learning [3][13]. So, it is the best practice to work with the existing curriculum to modify with new technologies.

A Research undertook the project ETHIKA for ethical education in pre-primary and primary schools for a sustainable and dialogic future to address the needs of pre-primary, primary, and lower secondary school students. The main outputs of the framework are educational materials and tools that are prepared with the previous User Needs Analysis (UNA) on students. Students learn important ethical principles and values, ethical reflection, awareness, responsibility, and intellectual capabilities. This framework also helps to build a classroom or school environment as an ethical community [14].

## III. PROBLEM ANALYSIS

There are some curriculums or frameworks like ETHIKA has developed to teach ethical and moral values to students. Educational institutions in Trinidad and Tobago emphasize moral and values education to develop student's strengths in problem-solving, aesthetic expression, technological competence, personal development. They propose some recommended topics for values education curriculum during the infant and childhood years such as Peace, Respect, Love, Responsibility, Happiness, Cooperation, Honesty, Humility, Tolerance, Unity, Freedom, Gratitude, Cleanliness, and Friendship [15]. Some international organizations like BBC also featured some ethical issues such as Introduction to Ethics, Abortion, Charity, Forced Marriage, Slavery, and Ethics of War [16]. But there is a common issue which is no/less involvement of technology to represent those theoretical lessons to the students. Students from pre-primary and primary may require more interactive lessons to learn about those theoretical concepts. AR technology may help to develop such interactive lessons where students can learn with fun and highly engaging will be ensured.

## IV. PROPOSED MODEL

We use the curriculum ETHIKA and select an educational resource called Tomato's Feeling from the Topic: Justice [17]. Initially, we design a system and develop a model where students play an AR game through a smartphone or tablet. We also need four flashcards containing pictures of tomatoes: a happy face, a sad face, an angry face, and a surprised face describes in Fig 2. We have to keep those cards separately on the table. A group of students plays the AR game with the guidance of a teacher. Some random question appears on the game interface related their feelings, students react to those questions by raising one of the four tomatoes. AR game shows an animated video based on the flashcard and records student's flashcard value. Then the teacher asks the student, "Why does this sentence make you feel this way?". AR Game gives better feedback or suggestion if the answer is not appropriate. This gameplay creates raw data for every student involved in the game and those data can be used to know the progress status of ethical development for students. Table 1 shows the key attributes of our proposed model.

TABLE I. KEY ATTRIBUTES FOR THE AR GAME

| Focusing Area | Justice |
|---|---|
| Schooling Level | Primary Level School |
| Class | Pre-primary |
| Age Range | 4 to 5+ |
| Participants | 5 to 10 Students |
| Materials and Tools | - Printed Flash Cards<br>- Smartphone/Tablet |
| Educational Methods | - Philosophy for and with children<br>- Holistic ethical learning |
| Key learning Outcomes | Encourage children to:<br>- Think about justice<br>- Realize what injustice is<br>- Realize the harmfulness of injustice |

TABLE II. QUESTIONS WITH PROBABLE ANSWERS

| Questions | Probable Answers |
|---|---|
| It is raining today. I forget to bring my umbrella so, I wet my dress when coming to school. | Sad/Angry |
| I had a chocolate but my friend ate it while I went to other room. | Sad/Angry |
| Look! What a beautiful cat it is. | Happy/Surprised |
| I gave a ball to a friend to mine but he threw it in the garbage. He did not like it. | Sad |
| My friend threw a toy out of the window and our teacher blamed me. | Angry |
| My brother was naughty and I didn't get the ice cream because of him. | Angry/Sad |
| I lost my favorite toy. | Sad |
| Today I cleaned up my room and my mother were very happy about it and praised me. | Happy |
| My father brought me my favorite football. | Happy |
| Wow, I just love my dress which my father gave me in my Birthday! | Surprised/Happy |

## V. SYSTEM DESIGN AND DEVELOPMENT

### A. Gameplay Steps

- Step 1: Teacher selects the particular class and student id/name from the game menu before student plays the game

- Step 2: Pick a random question from the Table 2, display it on phone's display, student understand the question and place the camera above the particular flash card

- Step 3: AR game detects the flashcard and shows a video related to the flash card's emotion

- Step 4: AR game give better feedback or suggestion if the answer is not appropriate

- Step 5: Repeat Step 2, 3 and 4 until all question finishes

- Step 6: Go to Step 1 if all student plays the game

*B. Gameplay Flowcharts*

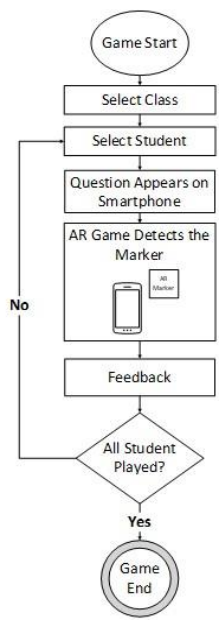

Fig. 1. Flowchart of gameplay

*C. Development*

We use Unity - a cross-platform game engine to develop our AR prototype game. Unity has options to develop games in both 2D and 3D and it offers various plugins and game resources for robust development [18]. Vuforia is an augmented reality software development kit (SDK) that enables the AR environment in the Unity game engine [19]. Vuforia creates AR markers for our tomato flashcards and Unity set conditions that which video is needed to show for a particular marker. We export our game to Android Package Kit (APK) which runs most of the Android smartphones.

## VI. PROTOTYPE AND RESULT

After compilation and build, our AR game successfully runs on an android based smartphone. Figure 3, shows that the AR game successfully detects different flashcards and shows the expected result based on the flashcards.

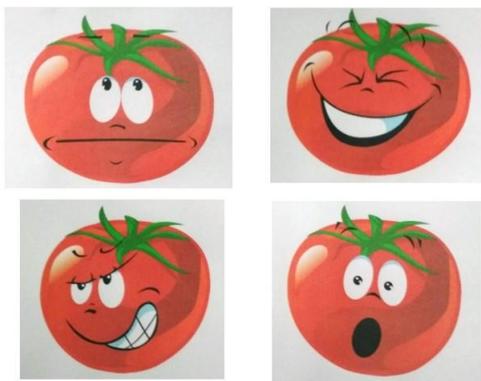

Fig. 2. Flashcards for the AR game

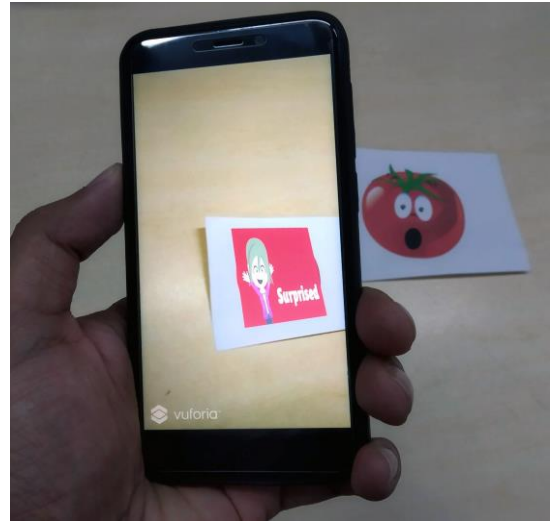

Fig. 3. Detecting flashcard by AR Game

## VII. PROTOTYPE AND RESULT

We develop an augmented reality-based mobile game model that enhances the engagement capability of the existing curriculum in the field of ethical education. The model simplifies holistic ethical learning and ethical philosophy with the help of augmented reality. Students also understand and learn about justice, realize what is injustice, and the harmfulness of injustice through the AR game in an interactive way.

So far, we have developed the system design model and a prototype for the AR game. We need to develop the full AR game in the future with the gamification technique. After developing the game, we need to test the game in the schools to justify the effectiveness of the game. We also need to design and develop the database for the game which may be needed to identify and track the progress in ethical education for a particular student.